# The Chandrasekhar's condition of the equilibrium and stability for a star in the nonextensive kinetic theory

Du Jiulin[*]

*Department of Physics, School of Science, Tianjin University, Tianjin 300072, China*

**Abstract**

The idea of Chandrasekhar's condition of the equilibrium and stability for a star is revisited in the nonextensive kinetic theory based on Tsallis entropy. A new analytical formula generalizing the Chandrasekhar's condition is derived by assuming that the stellar matter is kinetically described by the generalized Maxwell-Boltzmann distribution in Tsallis statistics. It is found that the maximum radiation pressure allowed at the center of a star of a given mass is dependent on the nonextensive parameter $q$. The Chandrasekhar's condition in the Maxwellian sense is recovered from the new condition in the case of $q=1$.

---

[*] Email Address: jiulindu@yahoo.com.cn

# 1. Introduction

Early in 1936, Professor Chandrasekhar presented a general theorem considered as equivalent to the condition for the stable existence of stars (Chandrasekhar,1936). The content of the theorem is the assertion that the actual pressure $P_c$ at the center of a star with a mass $M$ in the hydrostatic equilibrium must be intermediate between those at the centers of the two configurations of uniform density, one at a density equal to the mean density $\bar{\rho}$ of the star, and the other at a density equal to the density $\rho_c$ at the center; i.e., $P_c$ must satisfy the inequality,

$$\frac{1}{2}G\left(\frac{4\pi}{3}\right)^{1/3}\bar{\rho}^{4/3}M^{2/3} \leq P_c \leq \frac{1}{2}G\left(\frac{4\pi}{3}\right)^{1/3}\rho_c^{4/3}M^{2/3}, \tag{1}$$

Based on the right hand side of this inequality (1), the maximum radiation pressure at the center of a star of a given mass, $(1-\beta_*)$, was determined as the equilibrium and the stability condition for a star of a given mass. It is given with the mass $M$ of the star and the mean molecular weight $\mu$ by the equation (Chandrasekhar,1984),

$$\mu^2 M = 5.48\left(\frac{1-\beta_*}{\beta_*}\right)^{1/2} M_\odot, \tag{2}$$

where $M_\odot$ is the mass of the sun. From Eq.(1) it follows that for a star of solar mass with the mean molecular weight equal to 1, the radiation pressure at the center cannot exceed 3% of the total pressure, or the star will be unstable. As explained in the paper (Chandrasekhar,1984), such a result follows naturally from the kinetic theory in the Maxwellian sense, where the matter is assumed to satisfy the equation of state of an ideal gas.

On the other hand, a lot of recent studies on the statistical description for various physical systems, particularly for the systems endowed with long-range interactions, revealed that some extension of B-G statistical approach should be needed. Recently, the statistical mechanics based on the Tsallis nonextensive entropy has been developed as a very useful tool to describe the complex systems whose properties often cannot be exactly described by Boltzmann-Gibbs (B-G) statistical mechanics (Gell-Mann and Tsallis, 2004; Abe and Okamoto, 2001). This new statistical theory has been applied extensively to deal with varieties of interesting problems in the field of astrophysics where the systems are known to need the nonextensive statistical description due to the long-range nature of gravitational interactions (e.g., Lima, et al, 2002; Silva and Lima, 2005; Taruya and Sakagami, 2005, 2003; Leubner, 2004, 2005; Du, 2004a, 2004c, 2005; Hansen, et al, 2005). The Tsallis entropy is given (Tsallis, 1988) by

$$S_q = \frac{k}{1-q}\left(\sum_i p_i^q - 1\right), \qquad (3)$$

where $k$ is the Boltzmann constant, $p_i$ is the probability that the system under consideration is in its $i$th configuration, $q$ is a positive parameter whose deviation from unity is considered as describing the degree of nonextensivity of the system. In other words, (1-$q$) is a measure of the lack of extensivity of the system. The celebrated Boltzmann entropy is recovered from $S_q$ in the limit $q \to 1$. By using Eq.(3), if the system is composed of two subsystems numbered 1 and 2, then the total Tsallis entropy of the system is $S_q(1 \oplus 2) = S_q(1) + S_q(2) + (1-q)S_q(1)S_q(2)/k$. The extensivity is obtained when we take $q =1$.

Almost all the systems treated in statistical mechanics with B-G statistical mechanics have usually been extensive; this property holds for systems with short-range interparticle forces. When we deal with systems with long-rang interparticle forces such as Newtonian gravitational forces and Coulomb electric forces, where nonextensivity holds, B-G statistics may need to be generalized for the statistical description of such systems.

In this paper, we will revisit the idea of Chandrasekhar's condition of the equilibrium and stability for a star in the nonextensive kinetic theory based on Tsallis entropy, analyzing the effects of nonextensivity on the Chandrasekhar's condition and quantifying the role of nonextensivity in the equilibrium and stability of a star.

## 2. The nonextensive kinetic theory

The nonextensivity in the Chandrasekhar's condition Eq.(1) can be introduced through the equation of state of the ideal gas. In the nonextensive kinetic theory based on the Tsallis entropy, when we consider a self-gravitating system with particles interacting via the gravitational potential, the nonextensive velocity distribution function is given through the *q*-generalization of Maxwell-Boltzmann (M-B) distribution (Silva, et al,1998; Lima, et al, 2001; Du, 2004a, 2004d) by

$$f_q(\mathbf{r}, \mathbf{v}) = nB_q \left(\frac{m}{2\pi kT}\right)^{\frac{3}{2}} \left[1 - (1-q)\frac{m\mathbf{v}^2}{2kT}\right]^{\frac{1}{1-q}}, \qquad (4)$$

where *m* is the mass of each particle, *T* is the temperature, *n* is the particle number density, $B_q$ is the *q*-dependent normalized constant. The standard M-B distribution is recovered from Eq.(4) if we take *q* =1. This distribution has been applied to some interesting problems in astrophysics, such as the Jeans instability (Lima et al, 2002; Du, 2004b), the plasma oscillations (Lima et al, 2000), the negative heat capacity (Silva and Alcaniz, 2003), the solar wind intermittency (Leubner and Voros, 2004) and the dark matter (Hansen, et al, 2005; Leubner, 2005) etc.

We consider a cloud of ideal gas within the non-relativistic gravitational context. If *n* denotes the particle number density, in the view of kinetic theory, pressure of the gas is defined by $P_g = \frac{1}{3}nm<v^2>$ with $<v^2>$ the mean square velocity of the particle. When the nonextensive effect on the gas is considered in Tsallis statistics, it is introduced through a new expectation value of square velocity defined (Tsallis, et al, 1998) by

$$<v^2>_q = \frac{\int v^2 [f_q(\mathbf{r}, \mathbf{v})]^q d^3v}{\int [f_q(\mathbf{r}, \mathbf{v})]^q d^3v} \qquad (5)$$

The gas is now considered in the generalized M-B sense with the velocity distribution function Eq.(4). In this way, the *q* expectation value (5) for the square velocity of the particle is derived (Du, 2004b, Silva and Alcaniz, 2003) as

$$<v^2>_q = \frac{6}{5-3q}\frac{kT}{m}, \quad (0<q<\frac{5}{3}). \qquad (6)$$

It is clear that, as expected, the standard mean square velocity $<v^2> = 3kT/m$ is correctly recovered from the above equation if we take $q \to 1$. From Eq.(6) we can directly obtain the equation of state of an ideal gas in the nonextensive kinetic theory,

$$P_g = \frac{1}{3}nm<v^2>_q = \frac{2}{5-3q}nkT \qquad (7)$$

Eq.(7) may be written as the form familiar to astrophysics. Taking into account the particle mass $m = \mu\, m_H$ and the particle number density $n = \rho/\mu\, m_H$, where $\mu$ is the mean molecular weight and $m_H$ is the mass of the hydrogen atom, we have

$$P_g = \frac{2}{5-3q}\frac{k}{\mu\, m_H}\rho T \qquad (8)$$

The standard equation of state of an ideal gas is recovered perfectly from the above equations at the limit $q \to 1$.

## 3. The Chandrasekhar's condition

To gain a clear idea of effects of nonextensivity on the Chandrasekhar's condition, quantifying the role of nonextensivity in the equilibrium of a star, we consider Eq.(8) as the gas pressure, while the effect of nonextensivity on the Planck radiation law may be neglected. As usual, if the gas pressure contributes a fraction $\beta$ to the total pressure, using Eq.(8) and the radiation pressure, $P_r = aT^4/3$, where $a$ denotes the Stefan's constant, we may write the total pressure as

$$P = \frac{1}{\beta}\frac{2}{5-3q}\frac{k}{\mu\, m_H}\rho T = \frac{1}{1-\beta}\frac{1}{3}aT^4, \qquad (9)$$

Now, following the Chandrasekhar's line, from Eq.(9) we can derive the temperature,

$$T = \left[\frac{k}{\mu\, m_H}\frac{3}{a}\frac{2}{5-3q}\frac{1-\beta}{\beta}\right]^{\frac{1}{3}}\rho^{\frac{1}{3}}, \qquad (10)$$

Substituting back to Eq.(9) and eliminating the temperature $T$, we may express $P$ in terms of $\rho$, $\beta$ and $q$ by

$$P = \left[ \left( \frac{k}{\mu m_H} \right)^4 \frac{3}{a} \frac{1-\beta}{\beta^4} \right]^{\frac{1}{3}} \left( \frac{2}{5-3q} \right)^{\frac{4}{3}} \rho^{\frac{4}{3}}, \tag{11}$$

To bring out explicitly the role of the nonextensivity in the equilibrium of a star, we combine the right-hand side of the inequality (1) with $P$ given by Eq.(11) to yield, for the stable existence of stars, the condition,

$$M \geq \left( \frac{2}{5-3q} \right)^2 \left( \frac{6}{\pi} \right)^{\frac{1}{2}} G^{-\frac{3}{2}} \left( \frac{k}{\mu m_H} \right)^2 \left( \frac{3}{\alpha} \frac{1-\beta_c}{\beta_c^4} \right)^{\frac{1}{2}}, \tag{12}$$

where $\beta_c$ is the value of $\beta$ at the center. Then, replacing the Stefan's constant with $\alpha = 8\pi^5 k^4/15 h^3 c^3$, we have

$$\mu^2 M \left( \frac{5-3q}{2} \right)^2 \left( \frac{\beta_c^4}{1-\beta_c} \right)^{\frac{1}{2}} \geq 0.1873 \left( \frac{hc}{G} \right)^{\frac{3}{2}} \frac{1}{m_H^2} \tag{13}$$

On the right hand side of this inequality, the combination of nature constants can be given in units of the mass of the sun (Chandrasekhar, 1984), i.e., $(hc/G)^{3/2}/m_H^2 \approx 29.2$ $M\odot$. Thus, the condition (13) becomes

$$\mu^2 M \left( \frac{5-3q}{2} \right)^2 \left( \frac{\beta_c^4}{1-\beta_c} \right)^{\frac{1}{2}} \geq 5.48 \ M\odot \tag{14}$$

We may also obtain from this inequality an upper limit to $(1-\beta_c)$ for a star of a given mass so that

$$1-\beta_c \leq 1-\beta_{**}, \tag{15}$$

where the value of $(1-\beta_{**})$ is determined, but now related to the nonextensive parameter $q$, by the equation,

$$\mu^2 M \left( \frac{5-3q}{2} \right)^2 = 5.48 \left( \frac{1-\beta_{**}}{\beta_{**}} \right)^{1/2} M\odot \tag{16}$$

Thus the Chandrasekhar's condition Eq.(2) is generalized by Eq.(16) in the new kinetic theory when the nonextensivity is taken into consideration. As compared with Eq.(2) we find that

(a) $1-\beta_{**} > 1-\beta_*$ if $0 < q < 1$;

(b) $1-\beta_{**} < 1-\beta_{*}$ if $1< q <5/3$;

(c) $1-\beta_{**} = 1-\beta_{*}$ if $q =1$. (17)

**4. Conclusion and discussion**

From the above results we conclude that the maximum radiation pressure ($1-\beta_{**}$) allowed at the center of a star of a given mass is now dependent not only on its mass but also on its nonextensive parameter $q$. For example, if a star of solar mass is with the mean molecular weight equal to 1, the radiation pressure at the center can exceed 3% of the total pressure if $q<1$, for which the star was believed to be unstable while now may be still stable; on the contrary, the radiation pressure at the center may be less than 3% of the total pressure if $q >1$, for which the star was believed to be stable while now may become unstable. It is clear that Chandrasekhar's condition, Eq.(2), is recovered from Eq.(16) when in the case of $q=1$.

It is noticeable that how to understand the physical meaning of the nonextensive parameter $q$ plays a very important role in the applications of Tsallis statistics to the fields of astrophysics. However, in the light of present understanding, it is still an open problem. With the standard B-G statistics, the structure and stability of self-gravitating systems at statistical equilibrium are usually analyzed in terms of the maximization of a thermodynamic potential (the so-called mean field description). This thermodynamic approach leads to isothermal configurations (for example, self-gravitating isothermal gas spheres) that have been studied for long time in the stellar structure (Chandrasekhar,1942). It is well known that the stellar self-gravitating gases are usually in the hydrostatic equilibrium but not in the thermal equilibrium. The isothermal configurations only correspond to meta-stable states (locally convective mixing), not true equilibrium states. When the nonextensive effect is considered in the framework of Tsallis statistics, the equation of state of an idea gas can be written as $P_q = nkT_q$ with the physical temperature $T_q$ a variable (Abe, et al, 2001) that depends on the nonextensive parameter $q$. Comparing this equation with Eq.(7), we find $T_q = 2T/(5-3q)$. In this way, we agree with the explanation for $q$ of self-gravitating gases, i.e., $q=1$ represents an isothermal process of the gas, but $q \neq 1$ is corresponding to the non-isothermal one. For this point, we may introduce a relation between the nonextensive parameter $q$, the temperature gradient $\nabla T$ and the gravitational acceleration $\nabla \varphi$ that has been determined recently (Du, 2004a, 2004d) by

$$k\nabla T + (1-q)m\nabla \varphi = 0 \qquad (18)$$

This relation provides an physical interpretation for $q \neq 1$. $q$ is not unity if and only if $\nabla T$ is not equal to zero, i.e., the nonextensivity is closely related to the nonequilibrium degree of the system endowed with the long-range gravitating interactions. If applying Eq.(18) to the interior of a star, under spherical symmetry, we find

$$1-q = -\frac{k}{\mu m_H}\frac{dT}{dr} \bigg/ \frac{GM(r)}{r^2} \qquad (19)$$

There are still some interpretations for $q$ in some works. For example (Silva and Alcaniz, 2004), $q$ is related to the stellar polytrope index by n = 3/2 + 1/($q$-1) depending on the standard statistical averages or n =1/2+ 1/(1-$q$) depending on the normalized $q$ averages. In the limit n $\to \infty$, one has q $\to$ 1, corresponding to the isothermal distribution in an idea gas.

The different speeds of sound predicted by the tow statistics will cause a possibility to find the experimental evidence for the nonextensive effect. As a means of probing the interior structure and dynamics of a star with increasing precision, the helioseismology has provided the information about the square speed of sound through stellar interiors (Gough, et al, 1996). With this seismic observation, in principle, one might find the astronomical or experimental evidences for a value of $q$ different from unity (Du, 2006).

**Acknowledgements**

The author would like to thank the project of "985" Program of TJU of China for the financial support.

**References**

Abe S., Martinez S., Pennini F., Plastino A., 2001. Phys.Letts.A, **281**, 126.

Abe S., Okamoto Y., 2001, Nonextensive Statistical Mechanics and its Applications, Springer-Verlage, Berlin, Heidelberg.

Chandrasekhar S., 1936. MNRAS **96**, 644.

Chandrasekhar S., 1942. An Introduction to the Theory of Stellar Structure, Dover.

Chandrasekhar S., 1984. Rev.Mod.Phys. **56**,137.

Du J.L., 2004a. Europhys.Lett. **67**, 893.

Du J.L., 2004b. Phys.Lett. A **320**, 347.

Du J.L., 2004c. Physica A **335**, 107.

Du J.L., 2004d, Phys.Lett. A **329**, 262.

Du J.L., 2005. Centr. Euro.J. Phys. **3**, 376.

Du J.L., 2006. cond-mat/0602111.

Gell-Mann M., Tsallis C., 2004, Nonextensive Entropy-Interdisciplinary Applications, Oxford University Press, New York.

Gough D.O., Leibacher J.W., Scherrer P.H., Toomre J., 1996. Science **272**,1281.

Hansen S.H., Egli D., Hollenstein L., Salzmann C., 2005. New Astronomy **10**, 379.

Leubner M.P., 2004. ApJ **604**, 469.

Leubner M.P., 2005. ApJ **632**, L1.

M.P.Leubner, Z.Voros, 2004. ApJ **618**, 547.

Lima J.A.S., Silva R., Santos J., 2000. Phys.Rev.E **61**, 3260.

Lima J.A.S., Silva R., Santos J., 2002. A&A **396**, 309.

Lima J.A.S., Silva R., Plastino A.R., 2001. Phys.Rev.Lett. **86**, 2938.

Silva R., Alcaniz J.S., 2003. Phys.Lett. A **313**, 393.

Silva R., Alcaniz J.S., 2004. Physica A **314**, 208.

Silva R., Lima J.A.S., 2005. Phys. Rev. E **72**, 057101.

Silva R., Plastino A.R., Lima J.A.S., 1998. Phys.Lett. A **249**, 401.

Taruya A., Sakagami M., 2003. Phys. Rev. Lett. **90**, 181101.

Taruya A., Sakagami M., 2005. MNRAS **364**, 990.

Tsallis C., 1988, J.Stat.Phys. **52**, 479.

Tsallis C., Mendes R.S., Plastino A.R., 1998. Physica A **261**, 534.